\documentclass{osa-article}
\journal{oe}

\usepackage{physics}
\usepackage{subcaption}
\usepackage[normalem]{ulem}

\newcommand{\M}[1]{\mathbf{#1}}
\newcommand{\T}[1]{\mathrm{#1}}
\newcommand{\V}[1]{\boldsymbol{#1}}
\newcommand{\herm}{\T{H}}
\newcommand{\subto}{\T{s.t.}}

\newcommand{\rDep}{\left(\V{r} \right)}
\newcommand{\creg}{\T{c}}
\newcommand{\ureg}{\T{u}}

\renewcommand{\epsilon}{\varepsilon}

\begin{document}

\title{Fundamental bounds on the performance of monochromatic passive cloaks}

    \author{Lukas Jelinek,\authormark{1,*} Mats Gustafsson,\authormark{2} Miloslav Capek,\authormark{1} and Kurt Schab\authormark{3}}
    
    \address{\authormark{1}Department of Electromagnetic Field, Czech Technical University in Prague, Prague, Czech Republic \\
    \authormark{2}Department of Electrical and Information Technology, Lund University, Lund, Sweden \\
    \authormark{3}Department of Electrical and Computer Engineering, Santa Clara University, Santa Clara, CA, USA}
    
    \email{\authormark{*}lukas.jelinek@fel.cvut.cz} 

\begin{abstract}
Fundamental bounds on the performance of monochromatic scattering-cancellation and field-zeroing cloaks made of prescribed linear passive materials occupying a predefined design region are formulated by projecting field quantities onto a sub-sectional basis and applying quadratically constrained quadratic programming. Formulations are numerically tested revealing key physical trends as well as advantages and disadvantages between the two classes of cloaks. Results show that the use of low-loss materials with high dielectric contrast affords the highest potential for effective cloaking.
\end{abstract}

\section{Introduction}
Electromagnetic cloaks aim to minimize the interaction of a cloaked object with an incident field, resulting in reduced extinction cross section.  An ideal electromagnetic cloak is most commonly defined as a cover of arbitrary material composition that enforces zero scattered field outside of the cloak for an arbitrary illumination. The first proposals of such cloaks made of passive materials were described in~\cite{PendrySchurigSmith2006,Leonhardt2006} and were based on field transformation~\cite{Bladel1984,LeonhardtPhilbin2006} analogous to that performed to account for non-inertial frames of reference. An alternative point of view was later offered in~\cite{2008_Yaghjian_NJP,2008_Tretyakov_NJP} based on boundary-value formulation for material tensors and their direct synthesis, respectively. Following common terminology, such cloaks are referred to as ``isolated-object cloaks''. Soon after these developments, another cloaking strategy was introduced~\cite{2008_Li_PRL} in which the cloaked object lies on a reflecting plane and the objective of the cloak is to enforce the scattering properties identical to that of the reflecting plane alone. Such cloaks are commonly referred to as ``carpet cloaks'' and offer good cloaking performance with much simpler material distributions, making them attractive for practical realizations. These pioneering works together with experimental evidence of cloaking
\cite{SchurigMockJusticeEtAl2006} opened the broader topic of transformation optics~\cite{2014_Fleury_FERMAT}.

Despite many advances in cloak design, passive electromagnetic (EM) cloaks are severely limited in their performance. These limitations range from fundamental issues, such as the inability to screen static charges as dictated by Gauss's law, to more practical challenges, such as those arising from material dispersion, e.g., narrow bandwidth and loss. Severe restrictions are also imposed by the limited realizable values of refractive indices, which in turn dictates that the thickness of an effective cloak must be proportional to the thickness of the cloaked object itself~\cite{2011_Hashemi_PRA}. Typically, these detrimental effects are worsened as the electrical size of the cloaked object grows~\cite{2010_Hashemi_PRL,2012_Hashemi_PRA}. Fast-light cloaks using active media can potentially be used to increase the bandwidth and compensate for losses~\cite{2019_Tsakmakidis_NatComm,2021_Abdelrahman_NatComm}, but are out of the scope of this paper.

Limitations imposed on the frequency bandwidth of passive, linear and, time-invariant cloaks were studied in~\cite{2014_Monticone_PNFA}, with the conclusion that good cloaking performance within a finite frequency band must come at the expense of reduced performance over the remainder of frequency spectrum. This conclusion follows from the consideration that every passively cloaked object must exhibit increased extinction integrated over all frequencies when compared to an uncloaked object~\cite{Monticone+Alu2013,2010_Gustafsson_PRSA}.  Fundamental bounds on scattering from cloaked objects were derived in~\cite{2016_Monticone_Opca} based on field expansion into spherical harmonics with narrow-band excitation and in~\cite{2017_Cassier_JMP} based on analyticity for quasi-static cases. 

The fundamental bounds on cloaking performance mentioned above are relatively loose when compared to realized cloaks, i.e., they overestimate the performance achievable by practical systems. In contrast, bounds on general scattering and absorption behavior formulated as constrained optimization problems in volumetric contrast current densities or electric fields \cite{2020_Gustafsson_NJP,2020_Molesky_PRR, 2020_Venkataram_PRL, 2020_Kuang_PRL} lead to tighter, shape- and material-dependent bounds. Here, the terms shape- and material-dependent do not imply that exact knowledge of a particular cloak design is required before calculation of a bound, but rather that the bounding design volume is known along with some description of the materials that may occupy any portion of that design volume.  Despite the flexibility of these methods, optimization-based approaches have yet to be applied to the study of optimal cloaking.

In this paper, we adapt constrained optimization approaches to study the following question related to bounds on cloaking performance: \emph{Given a specified cloak material and operating frequency, what is the best possible passive cloaking device that can be formed out of that material within
a given volume surrounding a particular cloaked object?} Two formulations of this problem are studied using extinct power as a measure of cloaking performance (see Appendix~\ref{sec:PsVsPext} for further discussion regarding extinct and scattered power as metrics for cloaking performance). First, we examine ``scattering cancellation'' (SC) cloaks that minimize extinction by cancelling the field scattered by the cloaked object.  Because scattering cancellation depends on precise destructive interference of fields produced by the cloak and the cloaked object, the bounds imposed on these SC cloaks~\cite{2005_Alu_PRE,2008_Alu_JOA} necessarily depend on the cloaked object itself. As a shape-independent alternative, we also consider bounds on ``field zeroing'' (FZ) cloaks that shield the cloaked object from having any interaction with the outside world by producing fields which exactly cancel a specified excitation over a closed surface bounding the cloaked object. 

Throughout this work it is assumed that the cloaked object is made of arbitrary distributions of permittivity and permeability. In contrast, the material properties of the cloak are restricted. In one formulation, the cloak is, in line with existing experimental trials~\cite{SchurigMockJusticeEtAl2006,2014_Fleury_FERMAT}, assumed to be formed by an inhomogeneous distribution of macroscopic elements made of a single material (typically a lossy metal or lossy dielectric). Near-transparent host media or material supports are not considered since those do not induce qualitative changes in cloak performance. The second formulation allows for cloaks made of arbitrary material with greater thermal loss than a predefined value.  All materials are assumed to be passive and linear. Throughout this paper we adopt a time-harmonic convention~$\T{exp}\{-\T{i} \omega t\}$, where~$\omega$ is the angular frequency of the excitation.

\begin{figure}
    \centering
    \includegraphics[width=8cm]{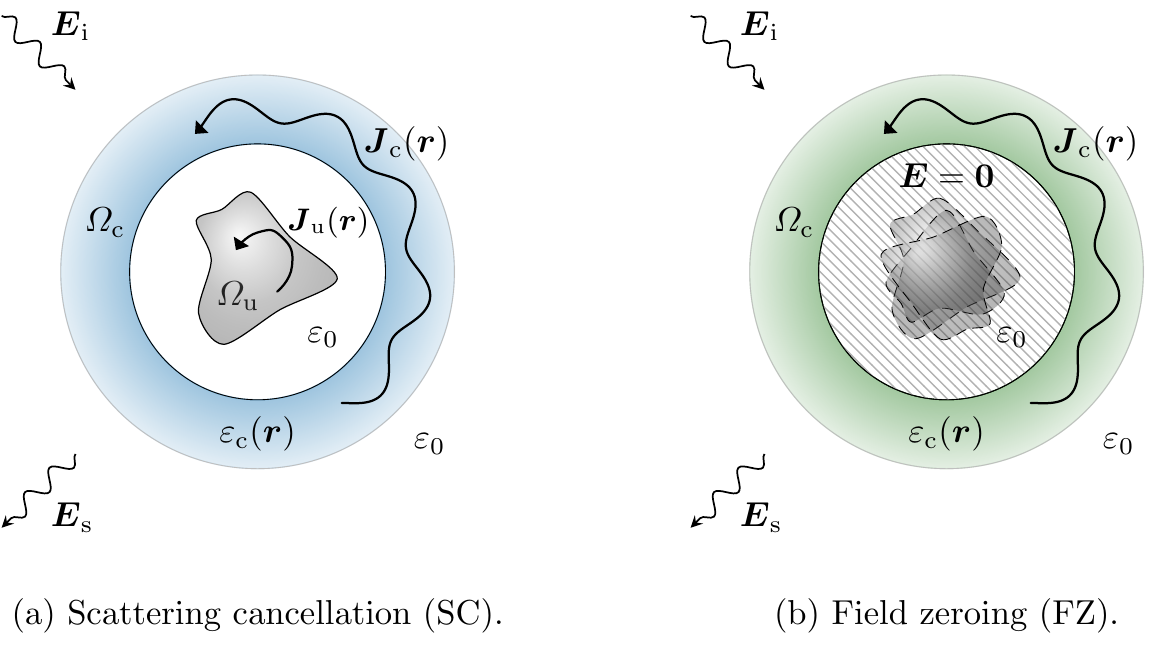}
    \caption{Schematic of cloaking setups described in this paper. The left panel shows an scattering-cancellation (SC) cloak where an incident wave impinges on a cloak residing in region~$\varOmega_\creg$ and a cloaked object residing in region~$\varOmega_\ureg$. When the fundamental bound on the performance of this system is formulated, the current density~$\V{J}_\creg \rDep$ is treated as an optimization variable obeying certain constraints. The right panel shows the setup of a field-zeroing (FZ) cloak. There, the cloak aims to provide zero field in the prescribed region (hatched), leading to the same cloaking performance for any arbitrarily shaped cloaked object within its interior (dashed outlines).}
    \label{fig:c-u-schem}
\end{figure}

\section{Scattering cancellation (SC) cloaks}
First, a theoretical framework for determining bounds on the optimal performance of SC cloaks is developed. Consider the situation depicted in the left panel of Fig.~\ref{fig:c-u-schem} where an incident field $\V{E}_\T{i} \rDep$ impinges on a structure composed of a cloak~$\varOmega_\creg$ and a cloaked material distribution $\varOmega_\ureg$ generating a  scattered~$\V{E}_\T{s} \rDep$ and total~$\V{E} \rDep = \V{E}_\T{i} \rDep + \V{E}_\T{s} \rDep$ field. The sub-index ``u'' denotes that the cloaked object $\varOmega_\ureg$ is ``uncontrollable'', i.e., that Maxwell's equations are strictly satisfied in region~$\varOmega_\ureg$, while the sub-index ``c'' denotes that a certain optimization scheme can ``control'' the field distribution in this region, potentially violating Maxwell's equations while still satisfying a set of relaxed constraints. Throughout this paper, the controlled variable is the contrast current distribution $\V{J}_\creg \rDep$ defined in region~$\varOmega_\creg$.

Given assumptions of linearity and time-harmonic steady state, the EM behavior of the system in Fig.~\ref{fig:c-u-schem} can be described by the field integral equation~\cite{1950_Lippmann_PR,ChewTongHu_IntegralEquationMethodsForElectromagneticAndElasticWaves}
\begin{equation}
    \mathcal{L}\V{J}(\V{r}) = \V{E}_\T{i}(\V{r}),\quad \V{r}\in \varOmega_\creg \cup \varOmega_\ureg,
    \label{eq:efie}
\end{equation}
where $\mathcal{L}$ is an operator related to the dyadic Green's function of the system, including material effects. By expanding the current into an appropriate basis $\left\{\V{\psi}_n \rDep \right\}$, relation \eqref{eq:efie} may be re-cast as a matrix equation~\cite{Harrington_FieldComputationByMoM}
\begin{equation}
    \M{ZI} = (\M{R - \T{i} \M{X} }) \M{I}= \M{V},
    \label{eq:ziv}
\end{equation}
where $\M{Z}$ is the impedance matrix, $\M{I}$ is the vector of expansion coefficients of the contrast current, and $\M{V}$ is a projection of the incident field onto the chosen basis. A great variety of scenarios can be described in this way using the method of moments, see~\cite{ChewTongHu_IntegralEquationMethodsForElectromagneticAndElasticWaves,Gibson_MoMinElectromagnetics} for a comprehensive list. Also note that a care should be taken when choosing basis functions~$\V{\psi}_n$ for a particular problem being studied, e.g., to properly model polarization charge on material boundaries. In the numerical examples shown in this work, the method of moments based on the volumetric electric field integral equation~\eqref{eq:efie} is used. Bodies of revolution are modeled with a Fourier-series expansion in the azimuthal coordinate and triangular elements in the other coordinates. Arbitrarily shaped bodies are treated with constant basis functions over a tetrahedral mesh. This treatment allows for the modeling of inhomogeneous material distributions.

In preparation for steps taken later in this paper, the impedance matrix is also decomposed to its imaginary part~$\M{X}$ and real part~$\M{R} = \M{R}_0 + \M{R}_\rho$, where~$\M{R}_0$ accounts for radiation and~$\M{R}_\rho$ accounts for thermal losses. The imaginary part $\M{X}$ may also be decomposed as $\M{X} = \M{X}_0 + \M{X}_\rho$.  As in the case of the real part $\M{R}$, the subscripts $0$ and $\rho$ indicate vacuum and material terms, respectively.  The material terms $\M{R}_\rho$ and $\M{X}_\rho$ are governed by the real and imaginary parts of the complex resistivity
\begin{equation}
\rho = \rho_\mathrm{r}+\mathrm{i}\rho_\mathrm{i} 
= \frac{\T{i}}{\omega\varepsilon_0}\chi^{-1},
\label{eq:resistivity}
\end{equation}
where $\varepsilon_0$ is the vacuum permittivity and $\chi$ is the material susceptibility, related to the material permittivity $\varepsilon$ and refractive index $n$ via
\begin{equation}
    \varepsilon = \varepsilon_0 n^2 = (1 + \chi)\varepsilon_0.
    \label{eq:eps-def}
\end{equation}
The real part of resistivity $\rho_\T{r}$ (or, to within a scaling factor, the imaginary part of susceptibility) is commonly used as a material figure of merit in other studies of physical bounds, e.g., \cite{MillerEtal_FundamentalLimitsToOpticalResponseInAbsorptiveSystems2016,2020_Gustafsson_NJP,2020_Schab_OPEX,2020_Molesky_PRR}, since it represents thermal loss mechanisms. Anisotropic materials may be modeled through the use of tensor, rather than scalar, quantities in \eqref{eq:resistivity} and \eqref{eq:eps-def}.  Anisotropic resistivity is specifically discussed further in Sec.~\ref{sec:results}.

If the basis $\left\{ \V{\psi}_n \rDep \right\}$ is sufficiently localized, i.e., no basis functions exist in both regions $\varOmega_\creg$ and $\varOmega_\ureg$ simultaneously, the system in \eqref{eq:ziv} may be partitioned as
\begin{equation}
\label{Eq:EFIE:cu}
\begin{bmatrix}
\M{Z}_{\creg\creg} & \M{Z}_{\creg\ureg}\\
\M{Z}_{\ureg\creg}& \M{Z}_{\ureg\ureg}
\end{bmatrix}
\begin{bmatrix}
\M{I}_\creg\\
\M{I}_\ureg
\end{bmatrix} = 
\begin{bmatrix}
\M{V}_\creg\\
\M{V}_\ureg
\end{bmatrix}.
\end{equation}

In line with the aim that only current in the controllable region~$\varOmega_\creg$, described by vector~$\M{I}_\creg$, is to be controlled, we eliminate the uncontrollable current by writing it in terms of the controllable current and excitation vector using the bottom row of \eqref{Eq:EFIE:cu} as
\begin{equation}
\label{Eq:EFIE:Schur}
\M{I}_\T{u} = \M{Z}_\T{uu}^{ - 1} \M{V}_\T{u} - \M{Z}_\T{uu}^{ - 1} \M{Z}_\T{uc} \M{I}_\T{c}.
\end{equation}
Upon substitution into the entire current vector $\M{I}$, this elimination leads to an affine transformation of variables (linear in the optimized variables) representing the entire current in terms of only the controllable current, i.e.,
\begin{equation}
    \label{Eq:transform}
\M{I} = \V{\alpha} + \sum\limits_n \V{\beta}_n x_n  = \V{\alpha} + \V{\beta} \M{x},
\end{equation}
where
\begin{equation}
    \V{\alpha} = \begin{bmatrix}
    \M{0} \\
    \M{Z}_\T{uu}^{-1}\M{V}_\T{u}
    \end{bmatrix}
    \quad \T{and} \quad
    \V{\beta} = \begin{bmatrix}
    \M{1} \\
    -\M{Z}_\T{uu}^{-1}\M{Z}_\T{uc}
    \end{bmatrix}
\end{equation} 
with $\M{1}$ being a square identity matrix,~$\V{\alpha}$ a fixed vector,~$\M{x}$ the new variables (in this case equivalent to the controllable current $\M{I}_\T{c}$), and with~$\V{\beta}_n$ being a set of new basis vectors forming the columns of the rectangular matrix~$\V{\beta}$. The important property of this choice of basis is that even when the total current~$\M{I}$ does not satisfy~\eqref{Eq:EFIE:cu}, the current in the uncontrollable region~$\M{I}_\ureg$ is strictly governed by the Maxwell's equations (i.e., the bottom row of~\eqref{Eq:EFIE:cu}) describing the incidence of the electric field generated by the controllable current~$- \M{Z}_{\ureg \creg} \M{I}_\creg$ and by the incident field~$\M{V}_\ureg$ on the uncontrollable region.  This procedure is analogous to the construction of Green's functions in the presence of scattering objects or boundaries~\cite{Tai_Dyadic_green_functions}, e.g., the use of image currents in problems involving infinite ground planes~\cite{Harrington_TimeHarmonicElmagField}. This property will now be used to form an optimization problem searching for bounds on the properties of all cloaks that can possibly be built within the region~$\varOmega_\creg$ from any arrangement of material with given complex permittivity.

Here we consider cloaks aiming to minimize the extinction cross section of the entire system (cloak and a cloaked object).  For any current $\M{I}$, the total extinct power is given by 
\begin{equation}
    P_\T{ext} = \frac{1}{2}\M{I}^\herm\M{R}\M{I},
\end{equation}
where~$^\herm$ denotes Hermitian conjugation.

Notice that if the current~$\M{I}_\creg$ would be required to satisfy~\eqref{Eq:EFIE:cu}, the total extinct power~$P_\T{ext}$ would show the effectiveness of the actual realization of a cloak built in the region~$\varOmega_\creg$. The shape of the best cloak is, nevertheless, not known. For that reason, the current~$\M{I}_\creg$ is allowed to vary freely as an optimization variable in order to set an absolute lower bound on~$P_\T{ext}$ from all possible cloak realizations fitting within the region~$\varOmega_\creg$ using its prescribed material properties. It is important to stress that the current~$\M{I}_\creg$ does not need to have all of its components non-zero if that is favorable for the optimized metric and that enforcing zero current density at a given point is equivalent to substituting the prescribed material by vacuum. With this in mind, the bound considered here automatically takes into account all possible distributions of vacuum and the prescribed material within the region~$\varOmega_\creg$. It is nevertheless important to notice that this does not yield the optimal material distribution in the controllable region, which is a task for much more computationally demanding topology optimization methods~\cite{2018_Molesky_NatP}.

The aforementioned bound can be formalized as a quadratically constrained quadratic program (QCQP)~\cite{NocedalWright_NumericalOptimization,BoydVandenberghe_ConvexOptimization} minimizing extinct power~$P_\T{ext}$ and taking~\eqref{Eq:transform} as a constraint. In such a formulation, however, the current~$\M{I}_\creg$ ignores Maxwell's equations~\eqref{Eq:EFIE:cu} and the bound thus formed is likely loose when compared to realized cloaks. To mitigate this deficiency, the relation~\eqref{Eq:EFIE:cu} is left multiplied by the conjugate current~$\M{I}^\herm$, forming a complex power balance~\cite{Harrington_TimeHarmonicElmagField,2020_Gustafsson_NJP}, which may be used to constrain the optimization over the current~$\M{I}_\creg$. The final optimization problem reads
\begin{equation}
\begin{aligned}
	& \min \limits_{\M{x}} && P_\T{ext} \\
	& \subto && \M{I}^\herm \M{Z} \M{I} = \M{I}^\herm \M{V} \\
	& && \M{I} = \boldsymbol{\alpha} + \boldsymbol{\beta} \M{x},
\end{aligned}  
\label{eq:QCQP2}
\end{equation}
see also Appendix~\ref{sec:OptimExplicit} for a more explicit form of this problem. The added requirement of power conservation enlarges the minimum attainable extinction~$P_\T{ext}$ when compared to an optimization omitting this constraint. However, any realized cloak satisfying \eqref{eq:ziv} must satisfy this form of power conservation and \eqref{eq:QCQP2} presents a tighter bound on the performance of all realizable cloaks. Addition of the power constraint also enables an alternative linear form of the objective extinct power, i.e.,~\mbox{$P_\T{ext} = \T{Re} \left\{ \M{I}^\herm \M{V} \right\} / 2$}.

The first constraint in \eqref{eq:QCQP2} may be decomposed into real and imaginary parts, corresponding to global conservation of real and reactive power, respectively.  The prescribed cloak material properties factor into each of these constraints in specific ways, with real resistivity (associated with losses) playing a role in real power conservation while imaginary resistivity (associated with reactance) impacting imaginary power conservation \cite{2020_Gustafsson_NJP}.  A relaxed form of \eqref{eq:QCQP2} may be constructed using only the real power constraint in order to form a bound on all possible cloaks constructed from materials with fixed losses and tunable reactances, like those achievable through reactive loading or metamaterial design.  The relaxed optimization problem is convex and reads
\begin{equation}
\begin{aligned}
	& \min \limits_{\M{x}} && \T{Re} \{\M{I}^\herm \M{V}\}  \\
	& \subto && \M{I}^\herm \M{R} \M{I} \leq \T{Re} \{\M{I}^\herm \M{V}\} \\
	& && \M{I} = \boldsymbol{\alpha} + \boldsymbol{\beta} \M{x},
\end{aligned}  
\label{eq:QCQP3}
\end{equation}
and represents a bound on the performance of any cloak constructed from inhomogeneous anisotropic material with real resistivity pointwise greater than or equal to that used in constructing the matrix $\M{R}$, see Appendix~\ref{sec:Rconst} for details.

\section{Field zeroing (FZ) cloaks}
Unlike the previously described SC cloaks, FZ cloaks present the same extinction behavior irrespective of the object being cloaked. Here, we develop an optimization problem analogous to \eqref{eq:QCQP2} to determine bounds on the optimal performance of FZ cloaks.

Within the setup depicted in the right panel of Fig.~\ref{fig:c-u-schem}, the only way to ensure that the system's scattering and absorption profiles are completely independent of the cloaked object is to enforce zero electric field throughout the cloaked object. Specifically, given an incident field $\V{E}_\T{i} \rDep$, the cloak must produce, through its induced current distribution~$\V{J}_\T{c}$, a scattered electric field~$\V{E}_\T{s} \left( \V{J}_\T{c} \right)$ satisfying
\begin{equation}
\label{eq:ConstZeroField1}
   \V{E}_\T{i} + \V{E}_\T{s} \left( \V{J}_\T{c} \right) = \M{0}, \quad \V{r} \in \varOmega_\ureg,
\end{equation}
where~$\varOmega_\ureg$ is any volume enclosing the cloaked object. Notice here that it is sufficient to satisfy~\eqref{eq:ConstZeroField1} only for tangential field components on the boundary of~$\varOmega_\ureg$. It may, however, be numerically advantageous to use condition~\eqref{eq:ConstZeroField1} as it stands.

In order to employ the previously presented matrix description of Maxwell's equations, we relax the above condition to require only the projection of the total fields upon the basis functions within the region $\varOmega_\ureg$ to be zero, rather than enforcing zero fields at all points of the region~$\varOmega_\ureg$. With this in mind, the weak form of \eqref{eq:ConstZeroField1} may be written as
\begin{equation}
\label{eq:cancelZIV}
\M{Z}_{\ureg\creg}\M{I}_\creg = \M{V}_\ureg,
\end{equation}
which enforces, in an averaged sense, that the electric field produced by the cloak current~$\M{I}_\creg$ cancels the incident field~$\M{V}_\ureg$ over the region $\varOmega_\ureg$.  By the invertibility of the matrix $\M{Z}_\T{\ureg\ureg}$ and the bottom row of \eqref{Eq:EFIE:cu}, this condition also implies zero current in the cloaked object, i.e., $\M{I}_\ureg = \M{0}$.   This is analogous to encasing the cloaked object in perfect electric conductor (PEC), though an optimal cloak could reduce the total extinction cross section below that of a PEC shield. Further, we assume that PEC objects might not be available and that conditions~\eqref{eq:ConstZeroField1} and~\eqref{eq:cancelZIV} must be enforced solely via the current distribution~$\V{J}_\T{c}$ within a prescribed non-PEC cloak material.

Like in the study of SC cloaks above, optimization problem over the cloak current $\M{I}_\creg$ is now formulated to determine bounds on minimally-extincting FZ cloaks satisfying the constraint~\eqref{eq:cancelZIV} which replaces the affine constraint in~\eqref{eq:QCQP2}. An alternative formulation of this problem involves transforming the relation~\eqref{eq:cancelZIV} into the form of~\eqref{Eq:transform}, making the optimization problem for the FZ cloak formally identical to \eqref{eq:QCQP2} and~\eqref{eq:QCQP3}. This formulation is detailed in the Appendix~\ref{Sec:AppZeroField}.

\section{Results}
\label{sec:results}
Using the two previously derived formulations of performance bounds on cloaking, numerical examples are presented to demonstrate the formulations' capabilities and to highlight a few of their key features. All results presented are for ``isolated-object cloaks'', though both formulations in this paper can be used to analyze bounds on ``carpet cloaks''~\cite{2008_Li_PRL}, provided appropriate changes to the dyadic Green's function, see \eqref{eq:efie}, are made to account for the presence of a reflective backing on which the cloak and cloaked object reside.

\begin{figure}
\centering
\includegraphics[]{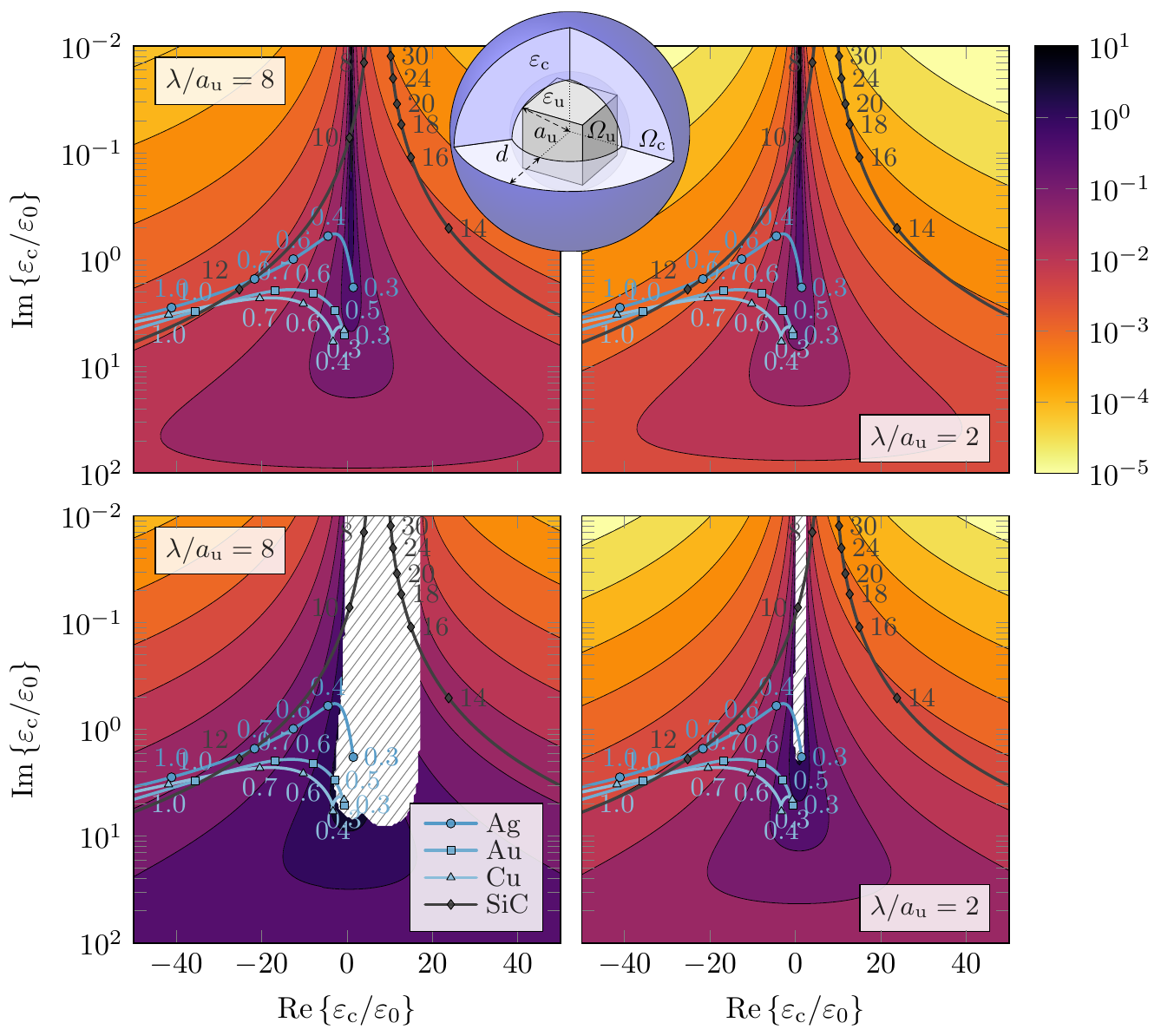}
\caption{Extinction efficiency~$P_{\T{ext}}/\big(S_0 \pi a_\T{u}^2 \big)$ of a SC cloak (top row) and FZ cloak (bottom row) at electric size (from left to right)~$ka_\T{u} = \{ \pi/4, \pi \}$ ($\lambda/a_\T{u} = \{8, 2 \}$) as a function of complex cloak permittivity $\varepsilon_\T{c}$.  The circumscribing sphere radii of the cloaked object (cube) and  zero-field region are~$a_\T{u}$. The thickness of the cloaking shell is~$d = a_\T{u}$. The optimization problem~\eqref{eq:QCQP2} with complex power conservation is used. The cloaked cube is made of a low-loss permitivity~$\epsilon_\T{r} = \{-5.7, -10 \}$ selected to maximize its uncloaked extinction efficiency~$P_{\T{ext},0}/\big(S_0 \pi a_\T{u}^2\big) \approx \{ 9.3, 1.7 \}$. The complex permittivities of silver, gold and copper~\cite{1998_Rakic_AO,Drude_Lorentz_at_file_exchange} in the range of wavelengths~$\lambda \in \left(0.3 \, \T{\mu m}, 1.1 \, \T{\mu m} \right)$ are also shown together with a complex permittivity of silicon carbide (SiC)~\cite{Palik_HandbookOfOpticalConstants} in the range of wavelengths~$\lambda \in \left(4 \, \T{\mu m}, 40 \, \T{\mu m} \right)$ containing resonance at wavelength~$\lambda \approx 13 \, \T{\mu m}$. Notice that resonance in permittivity of silicon carbide is not seen in the plot as due to its high absolute value it is outside the depicted range, presenting a poor material for cloaking in the vicinity of resonance frequency. For permittivites within the hatched region, the two optimization constraints in~\eqref{eq:QCQP2} cannot be satisfied simultaneously.}
\label{fig:FZ_sphere_ka_pi}
\end{figure}

The performance of a cloak is affected by the system's electrical size, the shape of the cloak and its material, the incident field distribution and, in the case of a SC cloak, the cloaked object itself. In reality all these effects are intertwined, making an exhaustive study of cloaking performance a complex, high-dimensional parametric sweep. In order to concisely present a few salient results, here we focus on numerical examples based on fixed incident field distributions, cloaked object shapes, and cloak materials commonly encountered in the literature.

In line with the majority of previous works on cloaking systems, a planewave excitation is used in this study to model the common scenario of a cloaked object illuminated by a far-field source. Furthermore, we restrict the shape of the cloaked object to that of a cube and a sphere. The cube is made of a low-loss permitivity~$\epsilon_\ureg / \epsilon_0 = \{-5.7, -10 \}$ selected to maximize its normalized extinct power (extinction efficiency)~$P_\T{ext}/\left(S_0 \pi a_\T{u}^2 \right) \approx  \{ 9.3, 1.7 \}$ at electrical size~$k_0 a_\T{u} = \{ \pi/4, \pi \}$, where~$k_0$ is a free-space wavenumber and~$a_\T{u}$ is the radius of a smallest sphere circumscribing the cube and~$S_0$ is the power flux of a planewave polarized along the edge of the cube and impinging normal to its face. The sphere is made of permittivity~$\epsilon_\ureg / \epsilon_0 = -2$ giving high extinction over large span of electrical sizes.

One of the important merits of the formulations presented in this paper are their independence on the particular shape of the cloak. Rather than analyzing the optimal performance of a specific geometry, the optimization procedure used here only requires the definition of a generic bounding region~$\varOmega_\creg$ in which the cloak resides, from which the best attainable performance is distilled among all possible realizations using the prescribed cloak material~$\varepsilon_\T{c}$. This allows us to choose a simple shape for the region~$\varOmega_\creg$ with almost no loss of generality, hence, the results presented here use a simple region~$\varOmega_\creg$ in a shape of a spherical shell of inner radius~$a_\T{u}$ and thickness~$d$. This is also the reason to normalize the extinct power to the cross-section~$\pi a_\ureg^2$.

The first example, based on optimization problem~\eqref{eq:QCQP2} with complex power conservation, is designed to examine the effect of a given cloak material on optimal cloaking performance. Following previous studies, it is expected that good cloaking performance requires a high contrast material~\cite{2011_Hashemi_PRA} which results in non-negligible electrical thickness of the cloaking shell, i.e., that it is the thickness of the cloak relative to the wavelength or skin depth inside the cloak material, not the physical thickness of the cloak, that impacts the performance. This is confirmed in Fig.~\ref{fig:FZ_sphere_ka_pi}, which shows the extinction efficiency of the entire cloaking system as a function of complex cloak permittivity. Comparing the optimized values with realized extinction efficiency of the cloaked object~($P_\T{ext}/\left(S_0 \pi a_\T{u}^2 \right) \approx  \{ 9.3, 1.7 \}$), it can be observed that the above expectations are valid for both SC and FZ systems and remain valid for different electric sizes as well.  The increase in cloaking performance at higher electrical sizes mostly results from the corresponding increase in electrical thickness of the cloak. Apart from electrical size, the need for high contrast material can also be, in line with the optimization procedure presented in this work, explained as providing more freedom to create ``material mixtures'' between the prescribed material and vacuum~\cite{2002_Milton_TheTheoryOfComposites} through current modulation inside the cloak. It is also important to notice that for low material contrast $\T{Re}\left\{\varepsilon_\creg / \varepsilon_0 \right\} \approx 1$ it is not possible to shield the object by the FZ cloak as illustrated by the hatched region. This issue becomes less significant at higher electrical sizes. The corresponding SC problem~\eqref{eq:QCQP2} is always feasible but the solution shows negligible improvement compared to the uncloaked object for low material contrast. 

\begin{figure}
\centering
\includegraphics[]{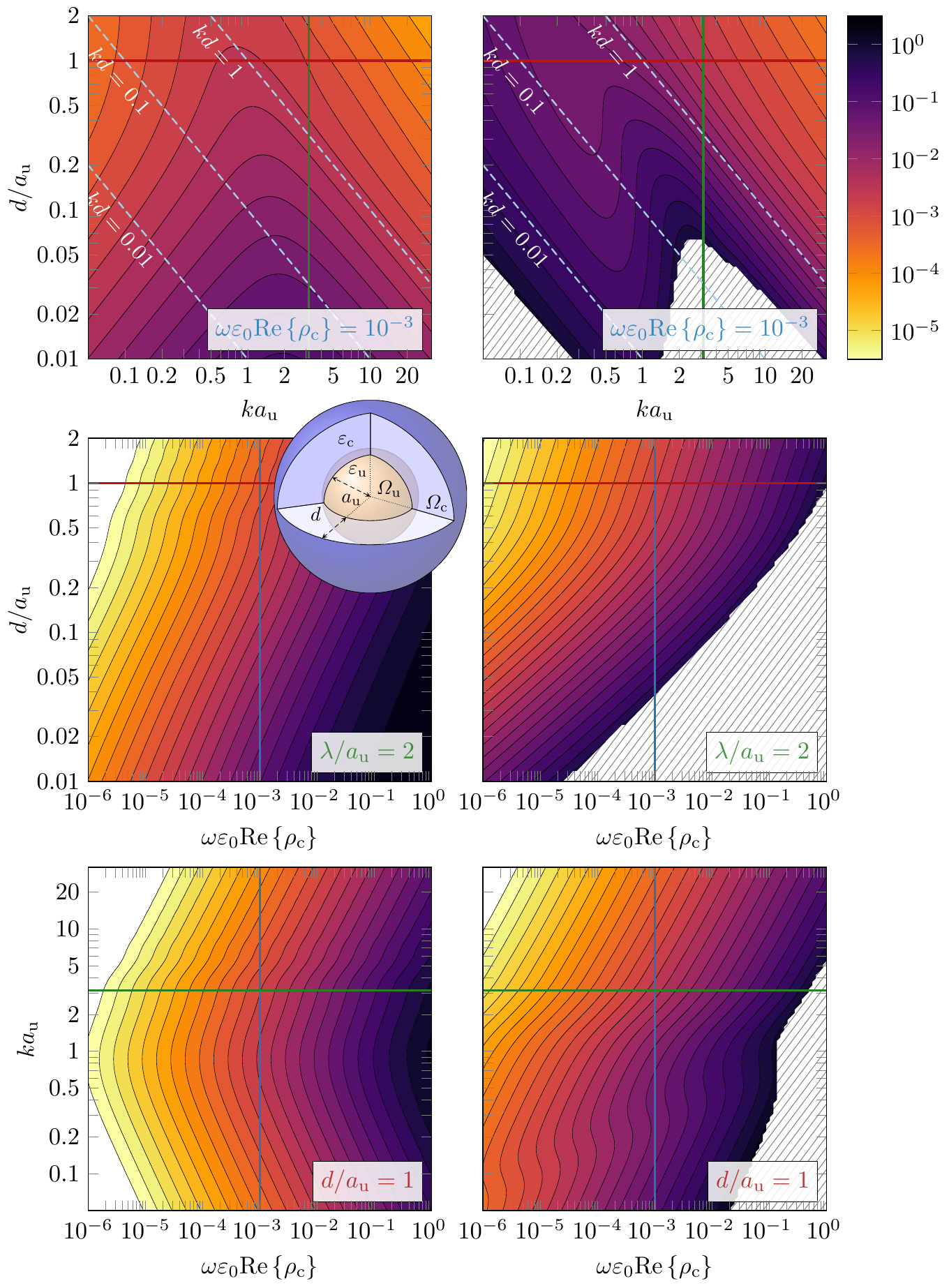}
\caption{Extinction efficiency~$ P_{\T{ext}}/\big(S_0 \pi a_\T{u}^2\big)$ of a SC cloak (left column) and FZ cloak (right column). The relaxed optimization problem~\eqref{eq:QCQP3} with only the real power conservation is used. In the first row, the cloak is assumed to be made of material with pointwise relation~$\omega \epsilon_0 \T{Re}\{\V{\rho}_\creg\} \succeq 10^{-3}\M{1}$, while this loss parameter varies in the second and third row. The circumscribing sphere radii of the cloaked object and  zero-field region are~$a_\T{u}$. The cloaked object (SC case) is a sphere made of a low-loss permittivity~$\epsilon_\ureg / \epsilon_0 = -2$. Parameters within the hatched region do not allow the optimization constraints to be satisfied.}
\label{fig:FZ_sphere_R}
\end{figure}

Better performance is achieved by the SC cloak due to its weakly constrained optimization problem, i.e., if it is advantageous for an optimal SC cloak to create zero fields over some region, it may do so, however an FZ cloak may not choose to relax its field zeroing constraint. Therefore, FZ cloaks can potentially meet, but never exceed, the performance of SC cloaks. When the contrast is high and losses $\T{Im}\{\epsilon\}$ are low, both cloaks, however, reduce extinction efficiency by several orders of magnitude as compared to the bare uncloaked object. The detrimental effect of increasing losses is severe. At a~$\T{Re} \{\epsilon_\creg / \epsilon_0 \}  = -40$ cut through Fig.~\ref{fig:FZ_sphere_ka_pi} there is approximately one order increase in extinction efficiency per one decade in imaginary part of permittivity. 

As a side note we mention that solving the optimization problem~\eqref{eq:QCQP2} is approximately 10 to 20 times computationally more costly than solving a scattering problem with the same discretization. In realistic scenarios, the solution of the optimization problem may however be of the same computational cost as simulating a realized cloaking structure as that might require finer discretization than that needed to discretize the optimal current density in the cloaking region.

A detailed analysis of Fig.~\ref{fig:FZ_sphere_ka_pi} reveals that normalized resistivity $\omega \epsilon_0 \rho_\creg = \T{i}(\epsilon_\creg / \epsilon_0-1)^{-1}$ better correlates with cloaking performance and that the bound on extinction efficiency is approximately proportional to $\omega \epsilon_0\T{Re} \{\rho_\creg\}$. This justifies the relaxed formulation~\eqref{eq:QCQP3} which is then used to determine bounds on the extinction efficiency for cloaks constructed by inhomogeneous and anisotropic materials. An example of this treatment is shown in Fig.~\ref{fig:FZ_sphere_R} for both the SC and FZ cloaks with several important cuts through the multidimensional design space. The cloaks are made of a material with an anisotropic resistivity tensor~$\T{Re}\{\V{\rho}_\creg\} \succeq \T{Re}\{\rho_\creg\}\M{1}$, which can be interpreted as having higher losses than an isotropic material with scalar resistivity $\T{Re}\{\rho_\creg\}$. The SC cloak assumes that the cloaked object is a dielectric sphere with $\varepsilon_\ureg / \varepsilon_0 =-2$ and radius $a_{\T{u}}$. 

The SC and FZ bounds are similar for electrically thick cloaks with $kd>1$, where the cloak is sufficiently thick to shield the object. They differ for electrically thinner cloaks where the SC bound depends weakly on the thickness but the FZ bound is close to unity before becoming infeasible for $kd<0.01$. The bounds are monotonically decreasing with decreasing ratio $d/a_\T{u}$, reinforcing the intuitive fact that it is always advantageous to be able to use a larger region to construct the cloak. The dependence for fixed ratios $d/a_\T{u}$ is more complex and has a peak around $ka_\T{u}=1$ for SC cloaks and a double peak for FZ cloaks. Dipolar interaction dominates for electrically small objects $ka_\T{u}\ll 1$ where SC cloaking is approximately solved by elimination of the core dipole moment through superposition with the dipole moment of the cloak~\cite[Chap.~7]{2002_Milton_TheTheoryOfComposites},~\cite{2005_Alu_PRE}. This behavior changes around $ka_\T{u}=1$, where higher-order modes start to contribute and the bound is dominated by the electrical thickness for $ka_\T{u}\gg1$. The SC and FZ bounds also depend on the minimum losses $\omega \epsilon_0\T{Re}\{\rho_\creg\}$. It is seen that the dependence is approximately proportional to the losses $\omega \epsilon_0\T{Re}\{\rho_\creg\}$, similarly to the cases in Fig.~\ref{fig:FZ_sphere_ka_pi}. 

\begin{figure}
    \centering
    \includegraphics[width=0.6\columnwidth]{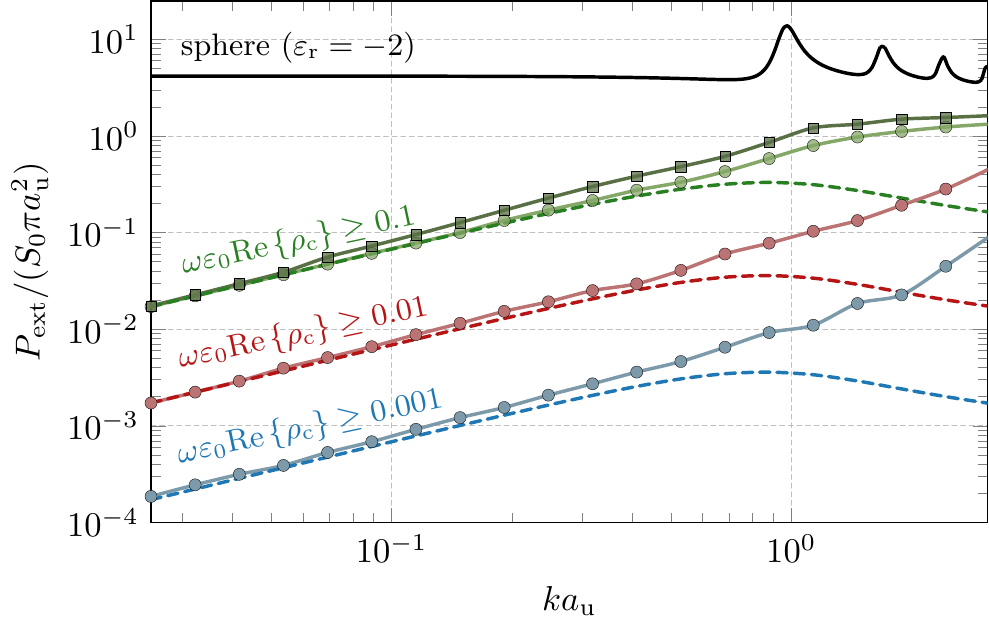}
    \caption{Comparison between the extinction efficiency~$ P_{\T{ext}}/\big(S_0 \pi a_\T{u}^2\big)$ for realized SC cloaks (solid curves with markers) with the lower bounds~\eqref{eq:QCQP3} (dashed curves). Cloaks and bounds are constructed from materials with losses $\omega \epsilon_0 \T{Re} \{\V{\rho}_\creg\} \succeq \left\{0.001, 0.01, 0.1\right\}\M{1}$ in a shell with thickness $d = a_\T{u}$ surrounding a dielectric core with permittivity $\epsilon_\ureg / \epsilon_0 = -2$. Realizations based on anisotropic and isotropic layers are shown using circular and square markers. The extinction efficiency of the sphere in isolation (no cloak) is shown by solid black curve.}
    \label{fig:realizations}
\end{figure}

An important aspect of fundamental bounds presented in this paper is their proximity to the performance of cloaking realizations. This comparison is shown in Fig.~\ref{fig:realizations} for relaxed SC bounds~\eqref{eq:QCQP3}, with ratio $d/a_\T{u}=1$ and minimum losses $\omega\epsilon_0\T{Re}\{\rho_\creg\}\in\{0.1,0.01,0.001\}$. Spherically symmetric cloaks based on multilayer structures with anisotropic and isotropic imaginary resistivity $\T{Im}\{\V{\rho}_\creg\}$ are considered as realized cloaks independent of the incident direction. Inverse design is used to determine the imaginary part in each layer for cloaks minimizing the extinction cross section. Decomposing the impedance matrix into its material and free-space parts $\M{Z}=\M{Z}_{\T{\rho}}(\V{\rho}_\creg)+\M{Z}_0$, rewrites the extinct power as
\begin{equation}
    \T{Re}\{\M{I}^{\herm}\M{V}\}
    =\M{V}^{\herm}(\M{Z}_{\T{\rho}}(\V{\rho}_\creg)+\M{Z}_0)^{-1}\M{V}
    \label{eq:InvDesign}
\end{equation}
which forms an unconstrained minimization problem over the imaginary components of $\V{\rho}_\creg$. This non-convex problem was solved using a Newton algorithm based on the explicit gradient and Hessian to find local minima of~\eqref{eq:InvDesign} combined with a global algorithm (here Monte Carlo and genetic algorithm were tested with similar results). Note that fixed losses $\omega\epsilon_0\T{Re}\{\rho_\creg\}$ do not prevent exclusion of specific regions since large amplitudes of $\omega\epsilon_0\T{Im}\{\rho_\creg\}$ effectively correspond to free space, i.e., $\epsilon =\epsilon_0+\T{i}/(\omega\rho_\creg)\to\epsilon_0$ as $|\omega\rho_\creg|\to\infty$. 

The comparison in Fig.~\ref{fig:realizations} shows that simple spherically multi-layered realizations are very close to the bound for electrically small objects. These synthesized realizations start to deviate from the bound around $ka_\T{u}=1$, where the bounds bend down but the performance of the realizations approach unity. Two reasons behind this discrepancy can be hypothesized.  First, it is possible that the selected combination of structural parameterization and inverse design algorithm are not capable of finding solutions approaching the bounds.  Alternative approaches to inverse design (including the use of asymmetric structural parameterizations) may lead to realized performance closer to the derived bounds.  On the other hand, there is no guarantee of tightness of the derived bounds, and it may be that, for electrically large systems, they are overly optimistic.  Further research may lead to still tighter bounds over broad frequency ranges.

\section{Conclusion}
In conclusion, the examples presented in this paper reinforce previous results suggesting that cloaking performance is maximized when low-loss and high-contrast materials are used.  However, the more fundamental contribution of this work are the underlying optimization formulations which provide bounds on cloaking performance in a variety of settings.  The examples presented here represent only a small number of cases to be studied.  We expect that these flexible methods of computing bounds on passive, monochromatic cloaking systems will provide benchmarks and set expectations for future cloaking devices, particularly those developed through the use of inverse design methodologies.

\appendix

\section{Extinct Power vs Scattered Power}
\label{sec:PsVsPext}
This section explains reasons for and outcomes of the decision to use extinct power, rather than scattered power~\cite{2009_Alu_PRL}, as an optimized metric. 

Following the methodology introduced in the main text, it is possible to reformulate all presented bounds using scattered power, rather than extinct power, with the only difference being a change of the optimized metric. The physical consequences of this change are nevertheless significant. The radiation part of impedance matrix~$\M{R}_0$ exhibits a large null space made up of non-radiating currents which can, through the optimization, be employed to generate vanishingly small scattered power. Minimizing scattered power in this manner, however, comes at the cost of creating a near-field of significant amplitude and along with it high absorption, a problematic result for cloaking devices aiming to have minimal impact on an incident field. This situation is analogous to impedance matching, which aims to remove reflections (scattering) while maximizing the power delivered to a load (absorption), i.e., matching makes the load very ``visible'' through a large, predominately absorption-based extinct power.  In contrast, when extinct power is minimized, large near-field scattering is no longer optimal as it would induce high loss and therefore high extinction. The optimization problem using minimum extinction as its objective is therefore better for describing the practical goals of cloaks, i.e., minimizing the scattered field over both near- and far-field domains. 

\section{Explicit Form of Optimization Problems}
\label{sec:OptimExplicit}

The main body of the paper considers optimization problems of the form
\begin{equation}
\begin{aligned}
	& \min \limits_{\M{x}} && P_\T{ext} \\
	& \subto && \M{I}^\herm \M{Z} \M{I} = \M{I}^\herm \M{V} \\
	& && \M{I} = \boldsymbol{\alpha} + \boldsymbol{\beta} \M{x}.
\end{aligned}  
\label{eq:QCQP2Suppl}
\end{equation}
This problem can be solved directly, however, it is computationally advantageous to remove the affine (linear dependence on the variable~$\M{x}$)  constraint~$\M{I} = \boldsymbol{\alpha} + \boldsymbol{\beta} \M{x}$ first and to separate the complex power constraint into two real valued constraints. To that point, a generic real-valued quadratic form is defined as
\begin{equation}
    \mathcal{Q}(\M{I},\M{A},\M{a},a_0) = \M{I}^\herm \M{A} \M{I} + \T{Re} \left\{ \M{I}^\herm \M{a} \right\} + a_0,
    \label{eq:QuadFormGeneric1}
\end{equation}
where~$\M{A}$ is a Hermitian matrix,~$\M{a}$ is a complex column vector, and~$a_0$ is a real constant. Using~\eqref{eq:QuadFormGeneric1}, the optimization problem~\eqref{eq:QCQP2Suppl} is rewritten as
\begin{equation}
\begin{aligned}
	& \min \limits_{\M{x}} && \mathcal{Q}(\M{I},\M{R}/2,\M{0},0) \\
	& \subto && \mathcal{Q}(\M{I},\M{R},-\M{V},0) = 0 \\
	& && \mathcal{Q}(\M{I},\M{X},-\T{i}\M{V},0) = 0 \\
	& && \M{I} = \boldsymbol{\alpha} + \boldsymbol{\beta} \M{x},
\end{aligned}  
\label{eq:QCQP2SupplQ}
\end{equation}
where the first complex power constraint in~\eqref{eq:QCQP2Suppl} was separated into its real (Hermitian) and imaginary (anti-Hermitian) parts in~\eqref{eq:QCQP2SupplQ}. The last step is an explicit substitution of the affine transformation~$\M{I} = \boldsymbol{\alpha} + \boldsymbol{\beta} \M{x}$ which forms the last constraint in~\eqref{eq:QCQP2Suppl} and~\eqref{eq:QCQP2SupplQ} and which makes the distinction between~SC and~FZ cloaks. Making this substitution transforms the quadratic form~$\mathcal{Q}(\M{I},\M{A},\M{a},a_0)$ into 
\begin{equation}
    \widetilde{\mathcal{Q}}(\M{x},\M{A},\M{a},a_0) = \M{x}^\herm \widetilde{\M{A}} \M{x} + \T{Re} \left\{ \M{x}^\herm \widetilde{\M{a}} \right\} + \widetilde{a_0},
    \label{eq:QuadFormGeneric2}
\end{equation}
with
\begin{equation}
\begin{aligned}
\M{\widetilde A} &= \boldsymbol{\beta}^\herm \M{A} \boldsymbol{\beta} \\
	\M{\widetilde a} &= \boldsymbol{\beta}^\herm \left( 2 \M{A}\boldsymbol{\alpha} + \M{a} \right) \\
\widetilde{a}_0 &= \T{Re} \left\{ \boldsymbol{\alpha }^\herm \left( \M{A} \boldsymbol{\alpha} + \M{a} \right) + a_0 \right\}.
\end{aligned}  
\end{equation}
Using this result, the original optimization problem~\eqref{eq:QCQP2Suppl} is written as
\begin{equation}
\begin{aligned}
	& \min \limits_{\M{x}} && \widetilde{\mathcal{Q}}(\M{x},\M{R}/2,\M{0},0) \\
	& \subto && \widetilde{\mathcal{Q}}(\M{x},\M{R},-\M{V},0) = 0 \\
	& && \widetilde{\mathcal{Q}}(\M{x},\M{X},-\T{i}\M{V},0) = 0,
\end{aligned}  
\label{eq:QCQP2SupplQTil}
\end{equation}
which is a form readily approachable using the tools of quadratically constrained quadratic programming, see~\cite[App. B]{2020_Gustafsson_NJP} for the general procedure employed in this work.

\section{Relaxation to Real Power Conservation}
\label{sec:Rconst}
The relaxed optimization problem~(8) for reciprocal anisotropic materials with loss (described by the real part of a resistivity tensor $\V{\rho}$) bounded from below by the loss associated with a real-valued isotropic resistivity $\rho$, i.e.,
\begin{equation}
     \rho(\V{r})\M{1}
     \preceq \T{Re}\,\{\V{\rho}(\V{r})\} 
     \label{eq:supp:rho-pointwise}
\end{equation}
is derived from~(7) in two steps. First the extinct power is expressed as $P_{\T{ext}}=\T{Re}\left\{\M{I}^{\herm}\M{V} \right\}/2$ and the reactive power is discarded in the constraint, leading to
\begin{equation}
\begin{aligned}
	& \min \limits_{\M{x}} && \T{Re} \{\M{I}^\herm \M{V}\}  \\
	& \subto && \M{I}^\herm \M{R}(\V{\rho}) \M{I} 
	= \T{Re} \{\M{I}^\herm \M{V}\} \\
	& && \M{I} = \boldsymbol{\alpha} + \boldsymbol{\beta} \M{x}.
\end{aligned}  
\label{eq:QCQPrelax1}
\end{equation}
In the second step, the equality in the first constraint is relaxed to an inequality and 
$\M{I}^\herm \M{R} \M{I}$ is further relaxed to
\begin{equation}
    \M{I}^\herm \M{R}(\rho) \M{I}
    \leq \M{I}^\herm \M{R}(\V{\rho}) \M{I},
\end{equation}
obeying \eqref{eq:supp:rho-pointwise}. The final optimization problem is convex and reads
\begin{equation}
\begin{aligned}
	& \min \limits_{\M{x}} && \T{Re} \{\M{I}^\herm \M{V}\}  \\
	& \subto && \M{I}^\herm \M{R}(\rho) \M{I} \leq \T{Re} \{\M{I}^\herm \M{V}\} \\
	& && \M{I} = \boldsymbol{\alpha} + \boldsymbol{\beta} \M{x}
\end{aligned}  
\label{eq:QCQPrelax2}
\end{equation}
Note that the relation~\eqref{Eq:EFIE:Schur} is independent of the material in the controlled region, which appears only in the matrix~$\M{Z}_\T{cc}$.  Thus, by solving a single instance of the problem in~\eqref{eq:QCQPrelax2}, it is possible to obtain bounds on cloaks constructed from \emph{any} anisotropic material obeying~\eqref{eq:supp:rho-pointwise}, with $\rho$ representing a reference minimum isotropic resistivity.

\section{Field-Zeroing Constraint}
\label{Sec:AppZeroField}
By virtue of the bottom line of the partitioned system 
\begin{equation}
\label{Eq:supp:EFIE:cu}
\begin{bmatrix}
\M{Z}_{\creg\creg} & \M{Z}_{\creg\ureg}\\
\M{Z}_{\ureg\creg}& \M{Z}_{\ureg\ureg}
\end{bmatrix}
\begin{bmatrix}
\M{I}_\creg\\
\M{I}_\ureg
\end{bmatrix} = 
\begin{bmatrix}
\M{V}_\creg\\
\M{V}_\ureg
\end{bmatrix},
\end{equation}
the field zeroing constraint 
\begin{equation}
\label{eq:supp:ConstZeroField1}
   \V{E}_\T{i} + \V{E}_\T{c} \left( \V{J}_\T{c} \right) = \M{0}, \quad \V{r} \in \partial \varOmega_\ureg,
\end{equation}
implies 
\begin{equation}
\label{eq:cancelZIVSupp}
\M{Z}_{\ureg\creg}\M{I}_\creg = \M{V}_\ureg,
\end{equation}
with $\M{I}_\T{u} = \M{0}$ and $\varOmega_\ureg$ describing the support of the currents described by $\M{I}_\T{u}$. In order to transform this constraint into the form of
\begin{equation}
    \label{Eq:transformSupp}
\M{I} = \V{\alpha} + \V{\beta} \M{x},
\end{equation}
one realizes that~\eqref{eq:cancelZIVSupp} is only satisfied when the number of degrees of freedom in the cloak current~$N_\creg = \T{dim}~\M{I}_\creg$ is larger than the number of degrees of freedom representing currents within the cloaked object~$N_\ureg = \T{dim}~\M{I}_\ureg$. When this is the case, the relation~\eqref{eq:cancelZIVSupp} is an under-determined system of equations ($N_\creg > N_\ureg$) which can be resolved via the singular value decomposition
\begin{equation}
    \label{eq:svdZ}
    \M{Z}_{\ureg\creg} = \M{U}_1 \M{\Sigma} \M{U}_2^\herm,
\end{equation}
where~$\M{U}_{1}$ and $\M{U}_{2}$ are unitary matrices and
\begin{equation}
\M{\Sigma} = \mqty[\V{\sigma} & \M{0} ]
\end{equation}
is a rectangular matrix containing singular values on the diagonal of matrix~$\V{\sigma}$. For later purposes it is also advantageous to partition the matrix~$\M{U}_2$ analogously to the matrix~$\M{\Sigma}$ as
\begin{equation}
\M{U}_2 = \mqty[\M{U}_2^\T{L} & \M{U}_2^\T{R} ]    
\end{equation}
where $\M{U}_2^\T{L}$ and $\M{U}_2^\T{R}$ are right eigenvectors associated with non-zero and zero singular values, respectively.

Substitution of~\eqref{eq:svdZ} into~\eqref{eq:cancelZIVSupp} gives
\begin{equation}
\label{eq:cancelZIVSupp2}
\M{I}_\creg = \M{U}_2^\T{L} \V{\sigma}^{-1} \M{U}_1^\herm \M{V}_\ureg + \M{U}_2^\T{R} \M{x},
\end{equation}
where~$\M{x}$ is arbitrary. The zero current condition~$\M{I}_\ureg = 0$ within the cloaked region then allows the zero-field constraint in \eqref{eq:supp:ConstZeroField1} to be written as~\eqref{Eq:transformSupp} with

\begin{equation}
\V{\alpha} = \mqty[\M{U}_2^\T{L} \V{\sigma}^{-1} \M{U}_1^\herm \M{V}_\ureg \\ \M{0} ]\quad \text{and} \quad
\V{\beta} = \mqty[\M{U}_2^\T{R} \\ \M{0} ].
\end{equation}

An important aspect of the FZ cloak is the dependence of its performance on the ratio~$N_\creg / N_\ureg$. Clearly, the higher the dimension~$N_\ureg$, the more difficult it is to satisfy the constraint in~\eqref{eq:cancelZIVSupp}, leading to a higher lower bound on extinction.

\section{Backmatter}

\begin{backmatter}
\bmsection{Funding}
We would like to acknowledge the financial support of this work by the Swedish Research Council and by the Czech Science Foundation under projects \mbox{No.~19-06049S} and \mbox{No.~21-19025M}.

\bmsection{Disclosures}
The authors declare no conflicts of interest.

\end{backmatter}

\bibliography{references}
\end{document}